\pdfoutput=1

\documentclass[preprint,showpacs,preprintnumbers,amsmath,amssymb,floatfix,endfloats*]{revtex4}
\usepackage{graphicx}
\usepackage{dcolumn}
\usepackage{bm}
\usepackage{amsfonts}
\usepackage{color}

\begin{document}

\title{Distributions of pore sizes and atomic densities in binary glasses
revealed by molecular dynamics simulations}

\author{Maxim A. Makeev$^{1}$ and Nikolai V. Priezjev$^{2}$}
\affiliation{$^{1}$Department of Chemistry, University of
Missouri-Columbia, Columbia, MO 65211}
\affiliation{$^{2}$Department of Mechanical and Materials
Engineering, Wright State University, Dayton, OH 45435}

\date[]{\protect\today}

\begin{abstract}
      We report on the results of a molecular dynamics simulation study of
binodal glassy systems, formed in the process of isochoric rapid quenching
from a high-temperature fluid phase. The transition to vitreous state
occurs due to concurrent spinodal decomposition and solidification of the
matter. The study is focused on topographies of the porous solid structures
and their dependence on temperature and average density. To quantify the
pore-size distributions, we put forth a scaling relation that provides a
robust data collapse in systems with high porosity. We also find that the
local density of glassy phases is broadly distributed, and, with increasing
average glass density, a distinct peak in the local density distribution is
displaced toward higher values.
\end{abstract}


\maketitle

\section{Introduction}
\label{intr}

 The glass transition is generally envisaged as a dynamic transition from
a thermodynamically equilibrium liquid state to a non-equilibrium
(glassy) state~\cite{r00,r01,r02}. The transition occurs in atomic,
molecular, and polymeric liquids (as well as in a wide variety of
other forms of matter), under the condition of rapid decrease in
temperature; that is, cooling of the liquid phase must be fast
enough in order to avoid material crystallization~\cite{r01,r02}.
Despite a long history of studies aimed at understanding the
vitreous state of matter, the phenomenon of glass transition still
remains to be one of the most intriguing unsolved problems in
condensed matter theory~\cite{r03,r04}. In the light of overwhelming
complexity of the glass problem, atomistic simulations of
judiciously chosen model systems can be of great assistance in
gaining a more elaborate understanding of various structural and
dynamical aspects pertained to glassy materials. One of the most
successful atomistic models, developed to date, is the Kob-Andersen
(KA) binary mixture model~\cite{r05}. Atomic simulations, performed
using the KA model, cover a wide range of topics and include - but
not limited to - the simulation studies of
Refs.\,\cite{r06,r07,r08,r09,r10,r11,r12}. The KA model have been
proven useful for gaining insight into topics ranging from
atomic-level structure of glasses to longer-range dynamical
correlation. Also, it was instrumental in revealing the existence of
spatio-temporal correlations in dynamical behavior of glassy
materials (i.e., dynamic heterogeneity~\cite{r13,r13a}).

\vskip 0.05in

  There exist a number of approaches to describe the glass transition
phenomenon. A somewhat simplified approach is based upon the concept of
free volume. Originally, the concept was introduced to describe the
glass transition in polymeric systems in Ref.\,\cite{r14}. The analytical
representation of free volume was later formulated by Doolittle~\cite{r15}.
Within the framework, developed in Ref.\,\cite{r16}, the glass transition
occurs as a result of decrease in free volume of the amorphous phase below
a certain value. This makes a possibility of an isochoric glass
transition in a glass-forming system seemingly counterintuitive, as
volume per atom in such system must decrease in a vicinity of the
transition. A possible explanation of vitrification at constant
volume is that the glass transition is accompanied by spinodal
decomposition~\cite{r17,r18,r19,r20,r21,r22}. Correspondingly, the
separation of the liquid phase into solid and void phases occurs;
the process of transition to a glassy phase is thus dynamically
concurrent with that of spinodal decomposition. The non-equilibrium
phenomenon of phase separation has been studied extensively in
binary mixtures {\it via} both theoretical (Monte Carlo simulations)
and experimental means~\cite{r23}. Within the mean-field theory
framework~\cite{r24}, the initial instability is defined by negative
chemical potential derivative with respect to concentration in
binary mixtures, and by negative pressure derivative with
respect to the local density in one-component fluids. On the other
hand, only a few computational studies of phenomena taking place
during rapid quenching of liquids below the glass transition at a
constant volume have been performed in the past.
\vskip 0.05in
     A classical molecular-dynamics simulation study of an isochoric
vitrification in model two-dimensional Lennard-Jones (LJ) binary
mixtures was reported in Ref.\,\cite{r25}; model polymeric systems
were treated by a similar computational approach in
Ref.\,\cite{r26}. Significant theoretical progress has been achieved
in Ref.\,\cite{r27}, where the glass transition was studied in model
systems comprised of liquid phase droplets of small volume fraction,
immersed in the glassy phase. The authors reported on a novel
mechanism of the dynamics of coarsening, which was found to be
driven by migration and eventual coalescence of liquid droplets in
the glassy phase. More recently, dynamics of the spinodal
decomposition in binary LJ mixtures, leading to vitrification, was
investigated in Refs.\,\cite{r28,r29}. Among many other findings,
the studies have provided important information on scaling
relations, underlying the temporal evolution of binodal systems
during a transition to a glassy state and reported on the effects of
simulation system size on the corresponding scaling behaviors.

\vskip 0.05in

To date, several experimental studies have been performed to
understand the constant volume
vitrification~\cite{r30,r31,r32,r33,r34,r35,r36,r37}. The works
include experimental studies of glass transition taking place in
confined environments. In Refs.\,\cite{r31,r32}, it was found that
an isochoric vitrification under nanometer-scale confinement is
realized in a close vicinity of the transition. More recent studies
have provided a rather convincing experimental evidence that the
glass transition in confined environments is an isochoric
process~\cite{r36,r37}. The major focus of the these
studies was centered on the issues related to dynamics of isochoric
transition to glassy state. In the present study, the
dynamical aspects are left beyond the scope. Rather, we concentrate
on the micro-structural as well as topographical properties of the
vitreous phases and make an attempt to bridge two extreme
cases of free volume distributions; the ones that are
characteristic for highly porous and dense glasses.
Correspondingly, we consider systems, where, at the upper bound, the
free volume is the largest one can achieve under the condition that
the quenched samples are solid. At the lower bound, the free volume
is close to the one characteristic for conventional vitreous
systems. Note that local regions in bulk glasses, where the free
volume differs from the average, can be regarded as structural
defects. Our approach provides interesting insights and connections
to experimental studies of bulk glasses.

\vskip 0.05in

The remainder of the paper is organized as follows. The {\it modus
operandi}, employed in this work, is described in the next section
(Sect.\,\ref{mod}). In Sect.\,\ref{res}, we present our simulation
results on the topography of porous glasses and temperature
dependence of the pore-size distributions. Also, we discuss the
local density distributions in the solid phases. Finally, in
Sect.\,\ref{sum}, we summarize our principal findings and draw
conclusions.

\section{Modus Operandi}
\label{mod}

In this work, the atomic systems are modeled as the KA binary
(80:20) mixture of particles~\cite{r05,r38} in a periodic box.
Within the KA model's framework, a pair of atoms $\alpha$, $\beta$ =
$\{A,B\}$ interact {\it via} the Lennard-Jones (LJ) potential of the
form:
\begin{align}
\label{eq1}
V_{\alpha\beta}(r) = 4\,\varepsilon_{\alpha\beta}[(\sigma_{\alpha
\beta}/r)^{12}-(\sigma_{\alpha\beta}/r)^{6}].
\end{align}
The parameters of the interatomic potential are set to
$\varepsilon_{AA} = 1.0$, $\varepsilon_{AB} = 1.5$,
$\varepsilon_{BB} = 0.5$, $\sigma_{AA} = 0.8$, $\sigma_{BB} = 0.88$,
and $m_{A} = m_{B}$~\cite{r38}. The cutoff radius of the potential
is fixed at $r_{c,\alpha\beta} = 2.5\,\sigma_ {\alpha\beta}$. The
units of length, mass, energy, and time are measured in $\sigma =
\sigma_{AA}$, $m = m_{A}$, $\varepsilon=\varepsilon_{AA}$, and
$\tau=\sigma\sqrt{m/\varepsilon}$. The equations of motion are
integrated using the Verlet algorithm with the MD time step of
$0.005 \,\tau$~\cite{r39}, and the temperature was controlled by
velocity rescaling. Note that this work adopts the simulation
methodology from Refs.\,\cite{r28,r29} in what related to the
dynamical evolution of the systems under consideration.
Correspondingly, all the behavioral features pertained to the
dynamics of the glass transition and the spinodal decomposition are
equivalent to those reported on in the above references. The initial
atomic configurations are prepared as follows. First, the systems of
$3\,\times10^{5}$ atoms were thoroughly equilibrated at the
temperature of $1.5\,\varepsilon/k_{B}$ during $3 \times10^{4}\tau$
at a constant volume. Five independent samples were prepared at each
density in the range $0.2\leq\rho\sigma^{3} \leq 1.0$. Thereby
equilibrated systems were then quenched to low temperatures; that
is, well below the glass transition temperature of $0.435\,
\varepsilon/k_{B}$~\cite{r28,r29}. The temperature of the glassy
phases was varied in the range from $0.02$ to
$0.20\,\varepsilon/k_{B}$, the increment being
$0.01\,\varepsilon/k_{B}$. At each fixed temperature, the atomic
systems were relaxed at the constant volume during the additional
time interval of $10^{4}\tau$ to form porous structures. It was
found that the final atomic configurations are stable and correspond
to solid phases at different densities. Moreover, the mobility of
atoms is suppressed; the atoms remain largely in their positions at
the time scales accessible to molecular dynamics simulations. In
Refs.\,\cite{r28,r29}, the authors found that the system-size
effects become negligible, when the number of atoms is no less than
$\simeq3\, \times10^{5}$. The choice of the system size, used in
this work, is based upon this finding.

\begin{figure}[!t]
\centerline{\includegraphics[width=10.2cm]{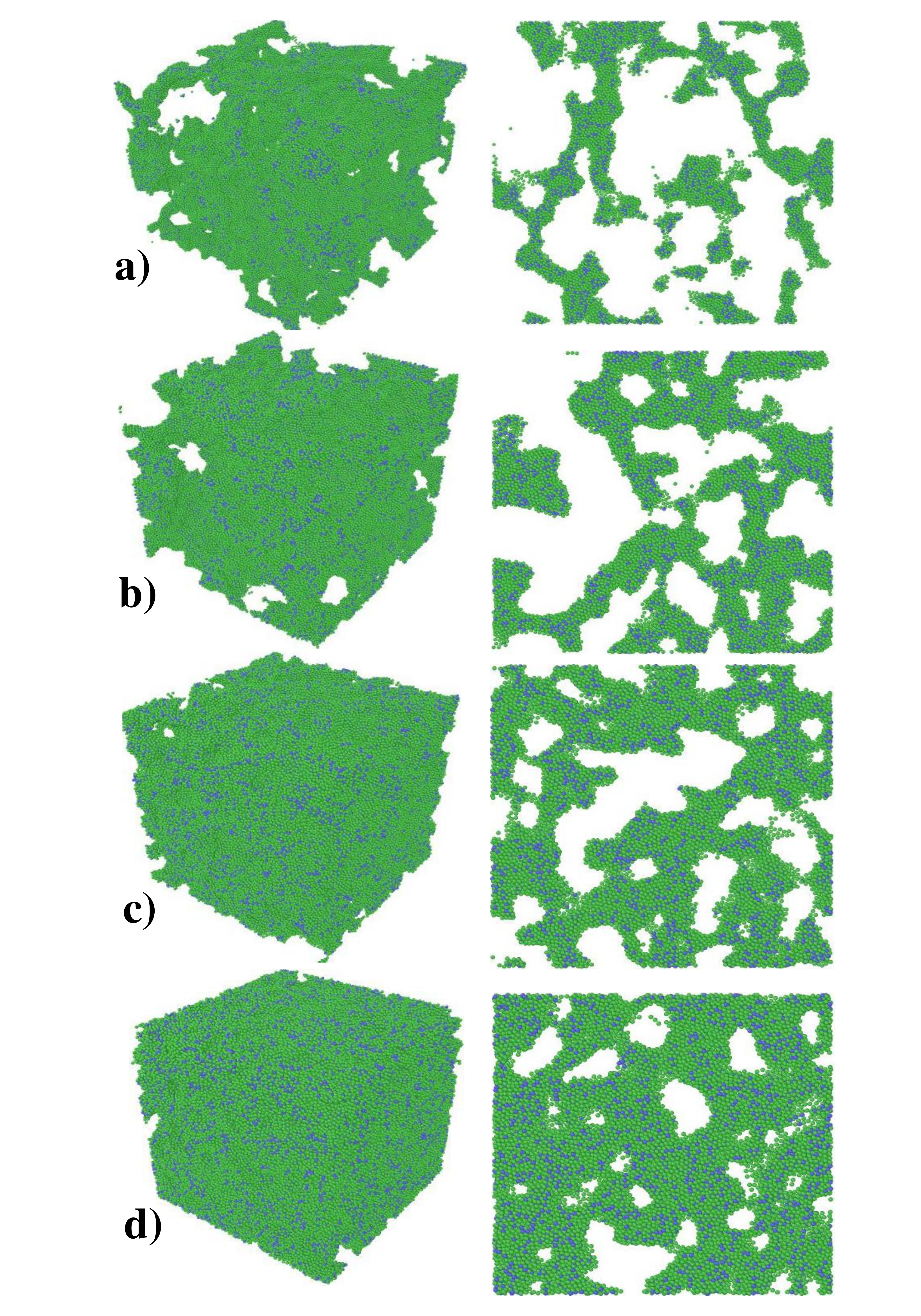}}
\caption{(Color online) Equilibrium instantaneous atomic configurations
of the glassy phases, computed at different values of $\rho\sigma^3$:
a) 0.3, b) 0.5, c) 0.7, and d) 0.9 (from top to bottom). All the
configurations are obtained at $T=0.05\,\varepsilon/k_{B}$. The left
panels show the full three-dimensional plots of the atomic configurations
and the right panels display slices of the central parts of the
simulation cell with thickness of $5\,\sigma$. Different colors mark
atomic types $A$ and $B$.}
%
\label{fig1}
\end{figure}

\section{Results}
\label{res}

\subsection{Notes on isochoric glass transition}
\label{glaT}

As made explicit above, the transition to glassy state at constant
volume occurs {\it via} concurrent spinodal decomposition into
material and void domains, and solidification in the systems
undergoing rapid quenching to a low-temperature state. In
Fig.\,\ref{fig1}, we show the representative examples of
the topographical patterns, obtained in our simulations study. The
major goal is to illustrate the larger-scale patterns in the glassy
samples. Two main factors in shaping topography of the porous
systems can be identified as the average density and temperature of
glassy phase. It was found that the topographical patterns vary
significantly with $\rho\sigma^3$. The free volume forms primarily
conglomerates of complex nanometer-sized shapes. Formation of
nanometer-scale channels running across the entire sample
also found to occur at low $\rho\sigma^3$. The channels possess
distinctly different characteristic length-scales, as compared to
having nanometer-scale (in all three dimensions) voids. Also
noteworthy is the observation that, in the whole range of average
density variation, the pores show no tendency to adopt even a
quasi-spherical shape. This observation can be of essence for
theoretical models, dealing with nano-porous continua. The
deviations from sphericity can be interpreted as resulting from
highly inhomogeneous and asymmetric tension in the systems. As
mentioned above, the degree of porosity in this work was varied in a
rather wide range. This allows for comparison of the present results
with findings reported in previous studies. While the
low-density (high porosity) systems have not been investigated in
sufficient details to have a comparative basis with experimental
works, highly dense systems have been a focus of experimental
studies, which makes a qualitative comparison feasible. Indeed, the
simulation results of our study show that the free volume
distribution at $\rho\sigma^{3}=0.9$ is that of a random quantity.
This is consistent with the recent experimental results of
Ref.\,\cite{r40}, where it was shown that the quenched-in free
volume in glasses is randomly distributed. Furthermore, the shape of
the distribution at the above density is similar to the experimental
results, obtained on the hole-radius density distribution in
polycarbonate and polystyrene~\cite{r41}. Further micro-structural
studies are needed to elucidate the nature of the glassy phase in
the binodal systems. A work in this direction is currently in
progress.
\vskip 0.05in
      Next, we turn attention to the process of the porous structure
formation. In Ref.\,\cite{r36}, the authors noted that ``...
molecular liquids confined in native nanopores form a glass at
constant volume under negative pressure". Furthermore, it was
observed that, in the region of positive pressures, the pressure
dependence of the glass transition temperature, $T_{g}$, can be
fitted to the Andersson-Andresson relation~\cite{r42}. The
extrapolation of the fit to the region of negative pressures was
shown to provide a rather accurate description of the experimentally
observed $T_{g}$ behavior. In our study, we also find that the
transition to the porous phase is driven by negative pressure. In
the past, a number of studies were devoted to the mechanism of void
formation under negative pressure. Thus, cavitation in LJ model
systems was studied in Ref.\,\cite{r43} using the Monte Carlo
method. It was found that there exists a critical density at which
the volume fraction of void vanishes, and this density coincides
with the minimum in the pressure {\it versus} density curve for
inherent structures (the Sastry curve) at a negative pressure. In a
subsequent study of Ref.\,\cite{r45}, the Sastry curves were used to
show that the properties of glass-forming mixtures depend on
softness of the interatomic potentials. In this respect, it should
be noted that our study employs the interaction potential with fixed
softness. Given all of the above, it is important to provide some
insights into the details of pressure variation with temperature of
the solid phase and as a function of $\rho\sigma^{3}$.

\begin{figure}[!t]
\centerline{\includegraphics[width=13.0cm]{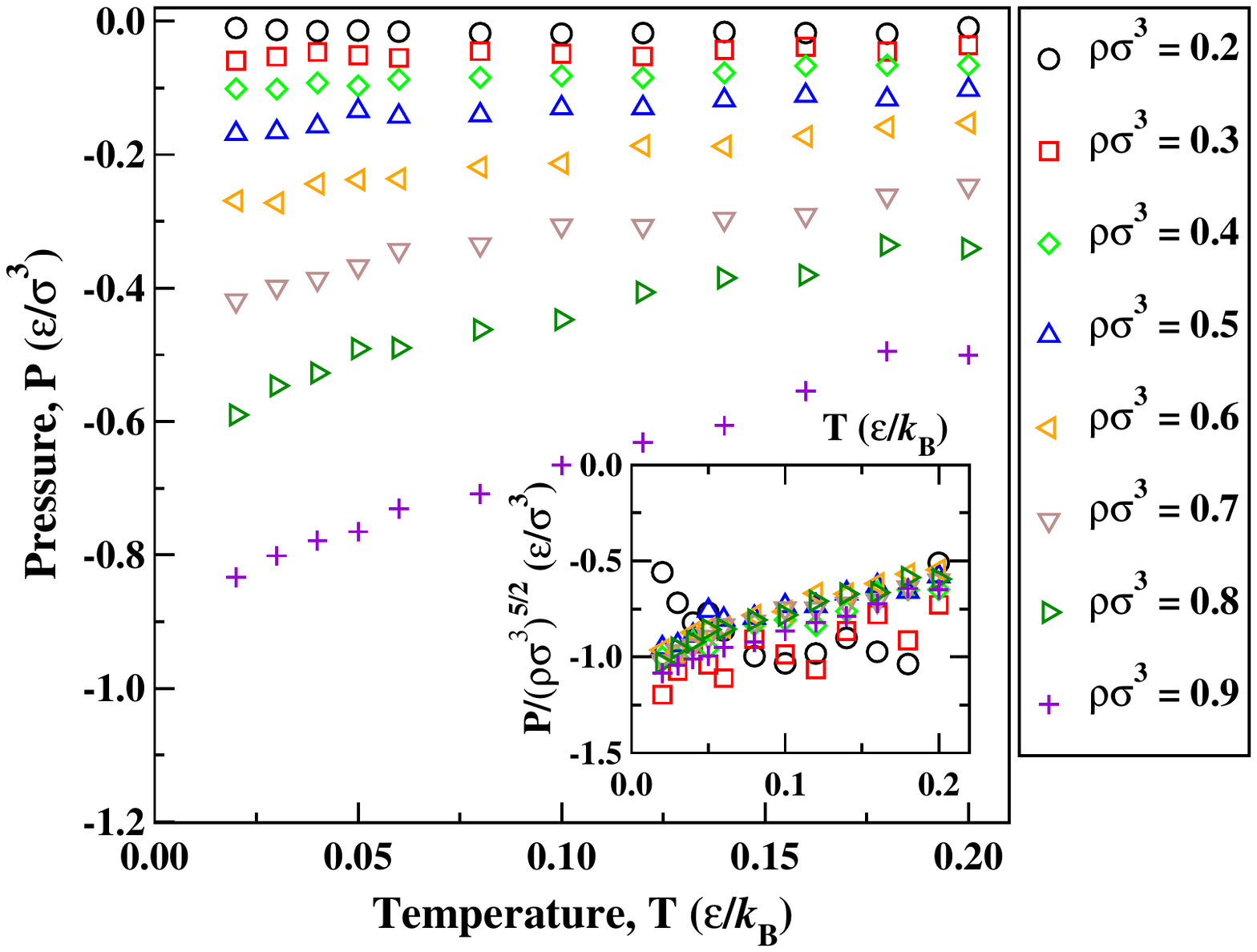}}
\caption{(Color online) The average pressure, $P$, in equilibrium
systems as a function of temperature, $T$, for the indicated
values of $\rho\sigma^3$. The inset shows the scaling collapse
using $P/T \sim \rho^{\alpha}$ relation. See text for details. }
%
\label{fig2}
\end{figure}

     In Figure\,\ref{fig2}, pressure versus temperature dependencies are
plotted for the model systems with different $\rho\sigma^{3}$. As follows
from Fig.\,\ref{fig2}, the transition to vitreous states and dynamical
evolution of the glassy systems take place under negative pressure, which
relaxes in the process of phase separation. At very low densities (see
$\rho\sigma^{3} = 0.2$ in Fig.\,\ref{fig2}), the pressure is relatively
small and its absolute value is an increasing function of temperature
when $T\lesssim 0.1\,\varepsilon/k_{B}$. As temperature increases above
$\simeq 0.1\,\varepsilon/k_{B}$, the pressure becomes a decreasing
function of temperature.

\vskip 0.05in

  The pressure variation as a function of temperature at densities $\rho
\sigma^{3} \ge 0.3$ is universal, and, depending on the temperature range,
three different regimes can be identified. As is evident in Fig.\,
\ref{fig2}, in this range of densities, the pressure increases monotonically
with temperature. However, the rate of increase at temperatures below
$\simeq 0.05\,\varepsilon/k_{B}$ differs considerably from those, observed
at intermediate temperatures, and those close to $\simeq0.20\,\varepsilon/
k_{B}$. Thus, at temperatures higher than $\simeq 0.07\,\varepsilon/k_{B}$,
the dependencies flatten out significantly. The decrease in negative
pressure is due to two important factors. First, the micro-structural
rearrangements are fully controlled by thermally activated processes.
However, the average density is the factor, which affects collective motion
and - in effect - shapes the pore topography. An increase in density
suppresses the mobility of atoms at a fixed temperature. Correspondingly, at
low temperatures, the relaxation is a strong function of temperature. In the
intermediate range of temperatures, between $\simeq 0.05\,\varepsilon/k_B$
and $0.15\,\varepsilon/k_{B}$, the rates of pressure variation are smaller
compared to the low-temperature range, yet a significant decrease in pressure
magnitude is observed at all densities studied. At low densities, this regime
corresponds to a nearly complete relaxation at any temperature (above $\simeq
0.15\,\varepsilon/k_{B}$). As expected, in the third region, a nearly flat
dependence is observed. Note that {\it qualitative features} of the
topographical patterns do not change with temperature. Also, in the whole
range of temperatures, the observed variations in pressure with temperature
are rather small, as compared to the effects due to the average density.
Indeed, the pressure changes by approximately an order of magnitude, while
the average density varies in the range $\rho\sigma^{3} \in [0.2, 0.9]$.

\vskip 0.05in

  In order to establish a more quantitative basis for estimating the effect
of density, in the inset to Fig.\,\ref{fig2}, we show the data collapse
obtained by using the scaling relation $P/T \sim \rho^{\alpha}$ with $\alpha
\simeq 5/2$. Given such a strong dependence, we conclude that pressure is
largely controlled by the {\it average density} in a wide range of $\rho
\sigma^{3}$. Also, pressure depends on topological properties of void domains.
One of the possible explanation for such a behavior is that it is associated
with surface energy. However, further studies of the near-surface effects
should be performed to uncover all the details of the observed behavior. Note
that the deviations from the scaling behavior are quite significant at
$\rho\sigma^{3} \leqslant 0.3$ and $\rho\sigma^{3} = 0.9$.

\begin{figure}[!t]
\centerline{\includegraphics[width=12.0 cm]{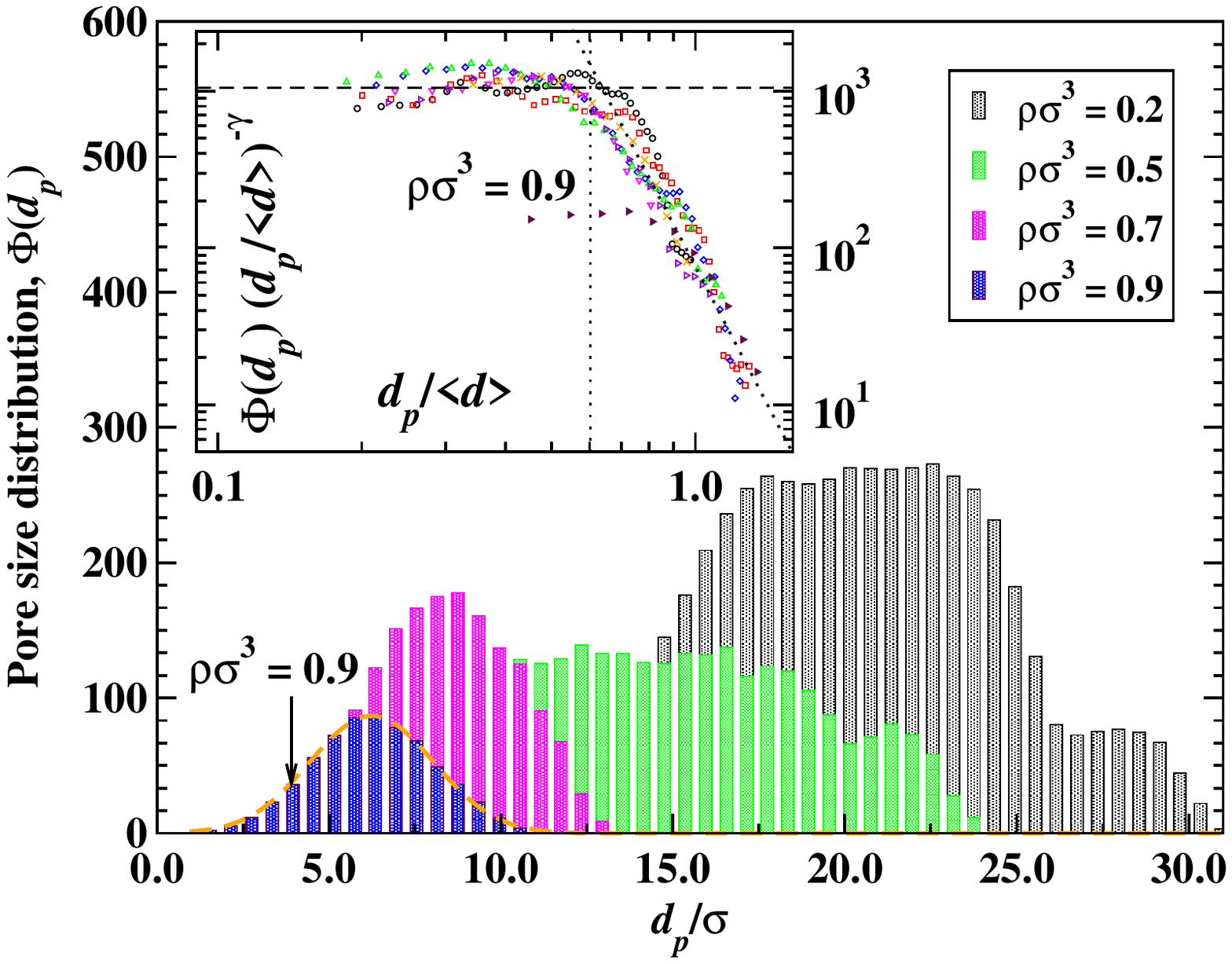}}
\caption{(Color online) The pore size distribution functions,
$\Phi(d_p)$, computed for systems with densities $\rho\sigma^3 \in
\left\{0.2, 0.5, 0.7, 0.9 \right\}$ and $T=0.05\,\varepsilon
/k_{B}$. In the analysis, the original bin size is fixed at
$\simeq0.05\,\sigma$. Subsequent averaging was performed within
$\simeq 0.5\,\sigma$; a reduced set of data points for each
$\rho\sigma^3$ is shown for clarity. The dashed orange curve is a
Gaussian fit to the $\rho\sigma^{3} = 0.9$ data. The inset shows the
data collapse for all densities according to Eq.\,(\ref{eq1}). The
same color code as in Fig.\,\ref{fig3}.}
%
\label{fig3}
\end{figure}

\subsection{Pore size distribution functions}
\label{psd}

  Further analysis involves pore-size, $d_p$, distribution (PSD) functions,
$\Phi$($d_{p}$). To quantify the topographical properties of the
ensembles of pores (see Fig.\,\ref{fig1} for illustration), we
employed the methods and the computer code developed in
Refs.\,\cite{r46,r47}. Fig.\,\ref{fig3} shows the behavior of the
PSD functions, computed at different values of $\rho\sigma^{3}$. As
can be observed in Fig.\,\ref{fig3}, in the region of small
pore-sizes, each PSD is an increasing function of $d_p$, for sizes
below a characteristic length-scale. Note that both the rate of
increase of the PSDs and their magnitudes are smallest at
$\rho\sigma^{3} = 0.2$. In general, both of these quantities are
increasing functions of $\rho\sigma^ {3}$. At larger $d_{p}$ values,
the PSDs flatten out. One more common feature of systems with
$\rho\sigma^{3} < 0.9$ is that they exhibit a peak at $d_{p}$ values
close to maximum pore-sizes. The above behavior is qualitatively
different from that of dense systems. In this respect, the cases
with $\rho \sigma^{3}=0.9$ and $1.0$ are representative examples.
Indeed, in dense systems, the PSDs profiles are close to Gaussian
(see the orange dashed curve in Fig.\,\ref{fig3}). In this case;
{\it i.e.}, the PSDs are fully described by the position of the peak
and the width of the distribution. As shown below, there exists a
universal scaling in the regime of small and intermediate
length-scales. To formulate the corresponding scaling law, let us
introduce the average pore diameter as: $\langle d
\rangle$=$\sum_{i=1}^{M}
d_{p(i)}^{2}\,n_{d_{p(i)}}/\sum_{i=1}^{M}d_{p(i)} \,n_{d_{p(i)}}$,
where $n_{d_{p(i)}}$ is the number of pores having discrete diameter
$d_{p(i)}$ and $M$ is the number of the discretization
points~\cite{r48}. Further, for the equilibrium continuous PSDs,
$\Phi(d_{p})$, we postulate the following scaling {\it ansatz}:
\begin{align}
\label{eq2}
\Phi(d_{p}) \sim (d_{p}/\langle d \rangle)^{\gamma} f(d_{p}/
\langle d \rangle).
\end{align}
The function $f(x)$ possess the following properties: a) $f(x\,\leq\,1)$
$\sim$ const; b) $f(0.5\,< x\,< 1.5)$ $\sim x^{-\gamma}$. We applied the
scaling form to the data obtained on all the considered herein systems.
The scaling collapse for all $\rho\sigma^{3}$ is shown in the inset to
Fig.\,\ref{fig3}. As can be observed in the figure, the data collapse is
quite convincing. The following findings are noteworthy. First, there
exists a universal exponent $\gamma \simeq 3$, which describes the data
for all densities up to $\rho\sigma^{3}=0.8$ inclusive. Second, the
average pore diameter is a strong function of porosity. By a numerical
analysis of the corresponding quantities, we found the following relation
between the average pore size and porosity: $\langle d \rangle\sim
p^{\delta}$, where $\delta$ is close to 2/3, as shown in Fig.\,\ref{fig3}.
In essence, this relation signifies the fact that the average pore size
is determined by the average surface area of the void space inside the
solid material. The maximum pore diameter, $d_{m}$, is defined by the
effective size of the available free space, which can be approximated as
$\lambda = p^{1/3}L$, where $L$ is the linear system size. The relation
holds for all densities, such that $\rho\sigma^{3}\leq 0.8$. Also, the
deviation from the scaling law, Eq.\,(\ref{eq1}), occurs at the length-
scales close to $d_{m}$. This indicates the existence of two independent
length-scales in the system: $\langle d \rangle$ and $d_{m}$. The effects
associated with the second length-scale manifest itself by the observed
deviations in the scaling behavior at $d_{p}/\langle d\rangle$ values above
1.5 [\,see the inset in Fig.\,\ref{fig3}\,].

\vskip 0.05in

\begin{figure}[!t]
\centerline{\includegraphics[width=13.0 cm]{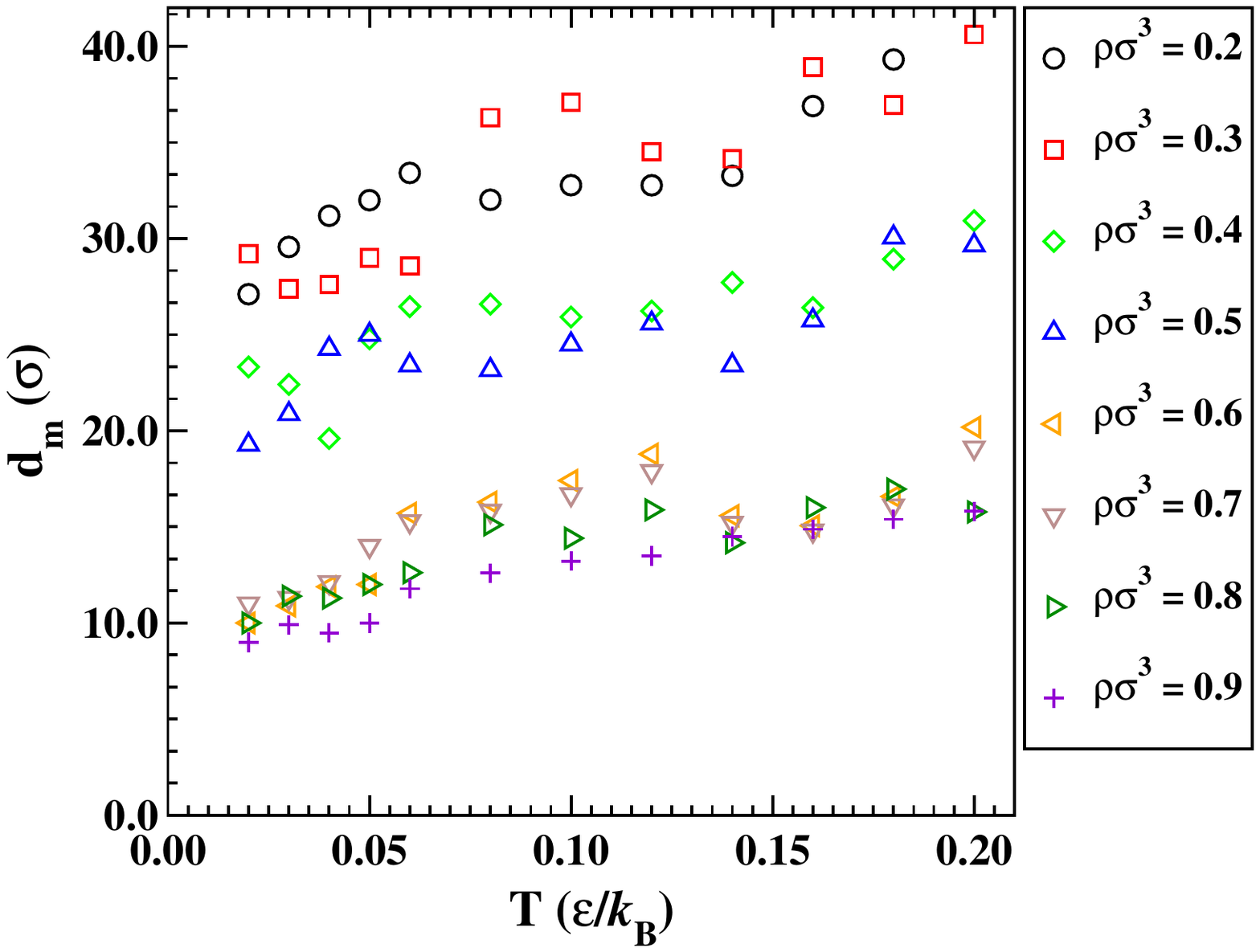}}
\caption{(Color online) The maximum pore diameter, $d_{m}$,
is plotted as a function of temperature for the tabulated
values of $\rho\sigma^3$.}
\label{fig04i}
\end{figure}

The behavior of the PSDs was studied in a temperature interval
between $0.02\,\varepsilon/k_{B}$ and $0.20\,\varepsilon/k_{B}$. We
found that the general shape of the PSD functions is preserved in
the whole range of temperature variation. At small pore diameters,
all the PSD functions follow a power-law behavior, with the
power-law exponent close to $\simeq3.0$. Further, in the range of
intermediate $d_{p}$ values, $\Phi$($d_{p}$) flattens out. Thus, the
scaling relation, Eq.\,(\ref{eq2}), holds in the whole temperature
range studied. Correspondingly, a variation in temperature changes
only the width and height of the PSDs, leaving the other features
intact. Note, however, that, in the proximity of the maximum pore
diameter, $d_{m}$, a peak is observed. We found that the magnitude
of the peak is nearly negligible, when temperature is less than
$\simeq0.05\,\varepsilon /k_{B}$. At higher temperatures the peak
magnitude increases, being $\simeq100\,\%$ greater than that of the
plateau region at $T=0.2\,\varepsilon/k_{B}$. As explained
above, the PSD functions can be regarded as piecewise combinations
of two distinctive part. The first part corresponds to small and
intermediate pore diameters and is described by Eq.\,(\ref{eq2}).
The second part corresponds to a separate peak in the vicinity of
$d_{m}$. In this regions, the scaling breaks down. Although, the
general shape of the PSDs is not affected by the variation in
temperature, the parameters of distributions vary substantially.
The maximum pore diameter versus temperature is plotted in
Fig.\,\ref{fig04i} at different values of $\rho\sigma^{3}$. As can
be observed in the figure, at every $\rho\sigma^{3}$ value, $d_m$ is
an increasing function of temperature, as expected. Moreover, the
relative maximum increments in $d_{m}$ due to temperature variations
are independent of $\rho\sigma^{3}$ and can roughly be estimated as
$\simeq30\,\%$. One curious feature of behavior is worth noting.
Three density ranges can be discriminated in Fig.\,\ref{fig04i}.
In each density range, the curves show remarkably similar behavior
in what related not only $d_{m}$ values, but also the specifics of
the functional dependence. From this behavior, one can infer that
there exist dynamical constraints on the pore evolution, such that
there is no continuous dependence of the pore-formation dynamics on
density.

\subsection{Local density distributions}
\label{lden}

\begin{figure*}
\includegraphics[width=15.0 cm]{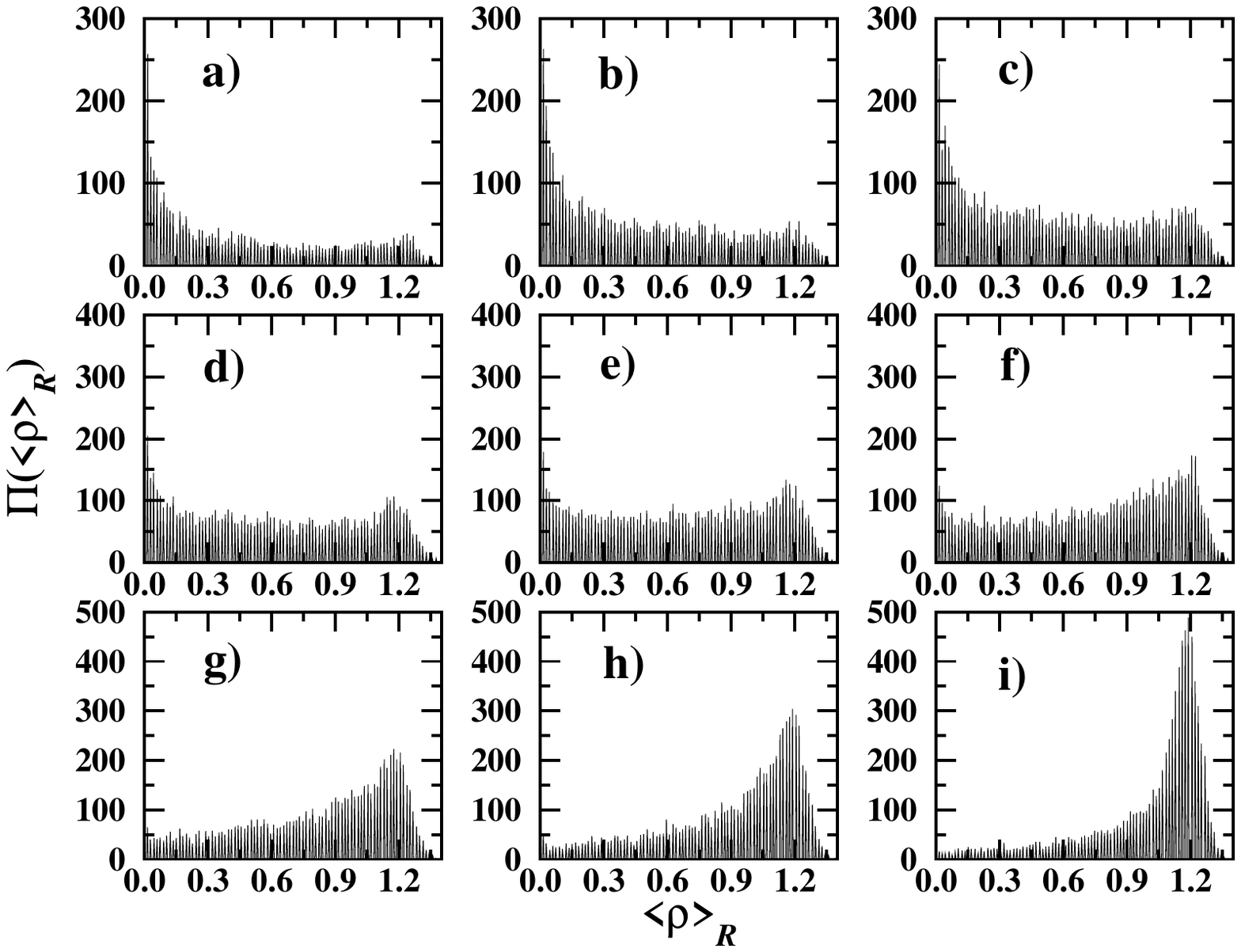}
\caption{On-lattice local density distribution functions, $\langle
\rho\rangle_{R}$, are shown for the normalized average density,
$\rho\sigma^3$, values: a) 0.2, b) 0.3, c) 0.4, d) 0.5, e) 0.6, f)
0.7, g) 0.8, h) 0.9, and i) 1.0. The distribution functions are
computed using the bin size of $\langle\rho\rangle_{R}^{max}$/400
and $T=0.05\,\varepsilon/k_{B}$; no averaging of any kind is
involved.}
\label{fig4n}
\end{figure*}

  In the preceding sections, we studied the properties of pores in
the glassy systems. In what follows, the focus is put on the solid
domains of the model systems under consideration. In particular, we
compute local densities in the solid-phase domains. Henceforth, the
local density is defined by a number of atoms located within a
predefined radial range centered on a site of the cubic lattice $L
\subset \mathbb{R}^{3}$. Correspondingly, for each lattice site, we
define a closed ball, $B_{R} = \{R \in \mathbb{R}^3: \sum_{j=1}^{3}
R_{j}^{2} \leq R_{0}^{2}\,\}$, where $R_{0} = |\vec{R}_{0}|$ is a
fixed rational number. Then, the average density is given by:
$\langle\rho\rangle = 1/V\sum _{i=1}^{N}\delta(\vec{r}_{i})=N/V=
\rho\sigma^{3}$. The on-site {\it local density} is computed as
$\langle\rho\rangle_{R} = 1/B_{R}\int_{B_{R}}dR^{3} \delta (\vec{r}
_{i}-\vec{R})$, where the integral is taken over $B_{R}$. Note that
the local density, $\langle\rho\rangle_{R}$, depends on $R_{0}$. In
our calculations, we used a fixed value of $R_{0} = 2.5\,\sigma$.
The rationale for choosing this value is based upon the behavior of
$\langle\rho\rangle_{R}(R_ {0}$) in dense (nearly void-free) glasses.
At this value of $R_{0}$ the local density becomes constant and thus
is equal to the {\it average density} of a homogeneous non-porous
glass. That is why, $\langle\rho\rangle_{R}$ can be regarded as {\it
a measure of deviation of the local density from the average density}
of homogeneous dense glass. In Fig.\,\ref{fig4n}, we plot the local
density distribution functions, $\Pi(\langle\rho\rangle_R)$, computed
at nine different average densities. As can be seen in the figure,
depending on $\rho\sigma^3$, three different types of behavior can be
discriminated (see lower, middle, and upper panels of the figure). In
the regime of small $\rho\sigma^3$ values (upper panel), the major
characteristic features are the following: a) a strong peak in the
vicinity of zero density; b) $\Pi(\langle\rho\rangle_R)$ is a
decreasing function of $\langle\rho\rangle_R$. This is not an
unexpected behavior for binodal systems with a large fraction of
pores. Indeed, pores and thin solid domains contribute to the peak
at zero and small value of $\langle\rho \rangle_{R}$. Note that the
surface to volume ratio in glassy phases is rather large in this regime
and therefore particles located in near-surface regions contribute
significantly to the behavior at intermediate $\langle\rho\rangle
_{R}$. The functional dependence with $\langle\rho\rangle_{R}$
resembles closely a continuous decay. Note, however, that there is a
peak in the region slightly above $\langle\rho\rangle_{R} = 1.2$. The
nature of the peak is discussed below.

 Lets us now turn the attention to the opposite case of dense, nearly
homogenous glass. The panels g), h) and i) in Fig.\,\ref{fig4n}
correspond to the large $\rho\sigma^3$ values. The behavior is, in a
sense, the opposite to the one, observed at small values of $\rho
\sigma^{3}$. Indeed, the highest peak is expected to be at a value
close to the average density of the solid phase. The plateau centered
at approximately $\rho\sigma^{3}=0.6$ is due to the near-surface
regions of the solid domains, where the numbers of atoms is close to
one half of the bulk. Once again, a peak at a value of $\langle\rho
\rangle_{R}$ close to $1.2$ is observed. In the region of small values
of $\langle\rho\rangle_{R}$, only a small deviation from zero can be
seen. Those are due to local fluctuations in free volumes that can be
regarded as microstructural defects. The peak in the intermediate
range of the local density are due to atoms in near-surface regions of
pores of small sizes. At the intermediate values of $\langle\rho
\rangle_{R}$, the patterns can be regarded as a superposition of the
two cases, discussed in the above. The solid domains are represented
by the peaks at the densities characteristic for large values of $\rho
\sigma^{3}$. The peaks near zero decrease in magnitude with the average
density, while, at intermediate values of $\langle\rho\rangle_{R}$,
the behavior is characteristic for bimodal systems with a large
surface-to-volume ratio.

\vskip 0.05in

\begin{figure}[!t]
\centerline{\includegraphics[width=12.0 cm]{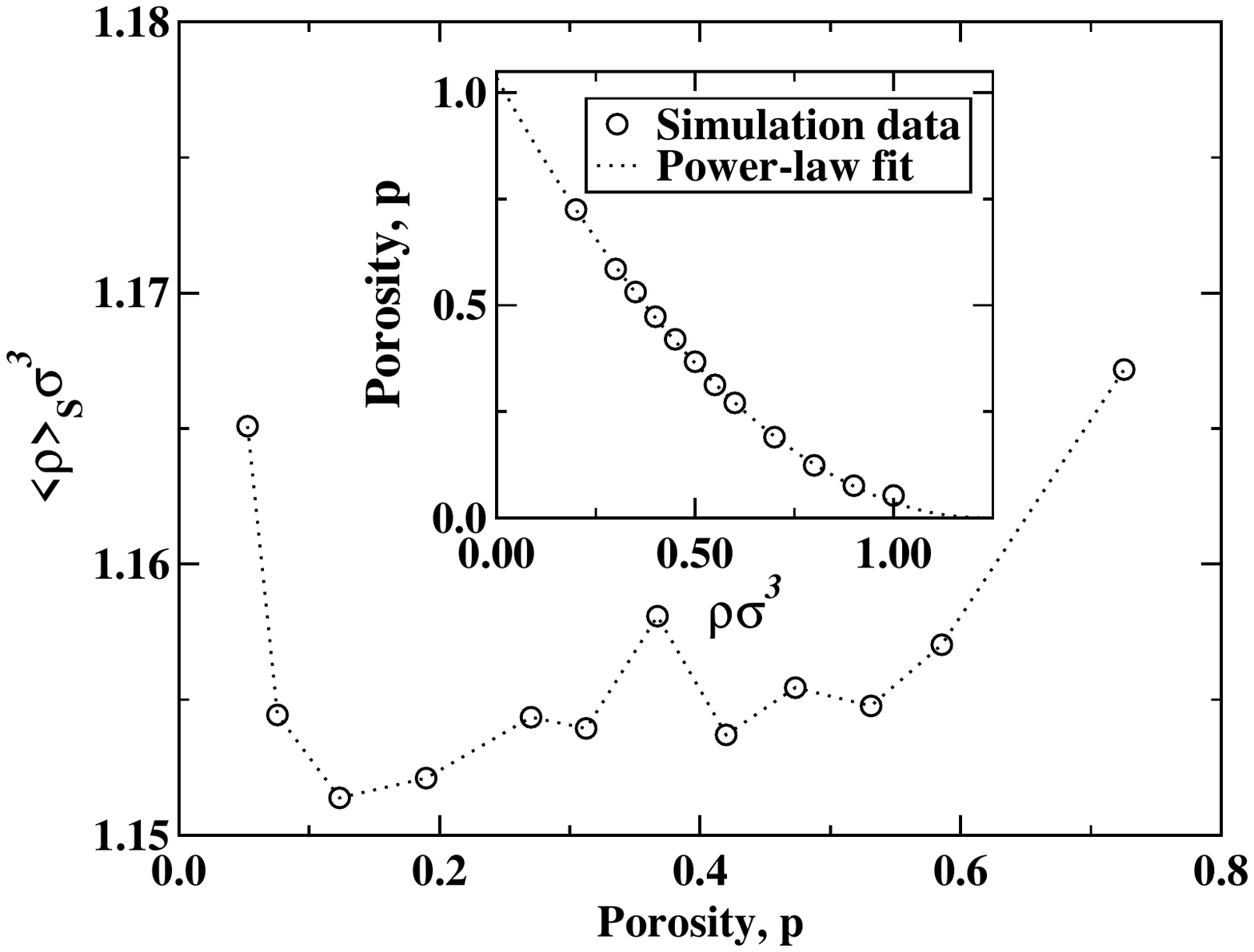}}
\caption{Average density of solid-phase domains,
$\langle\rho\rangle_{S}$, is plotted as a function of the porosity,
$p$. The inset shows the porosity versus $\rho\sigma^3$
dependence (open circles). The power-law fit to the simulation data
is indicated by the dashed curve. The fitting procedure is
discussed in the text. }
%
\label{fig6}
\end{figure}

The behavior of $\Pi(\langle\rho\rangle_R)$ functions in the regions
close to the maximum density requires an additional analysis. Those
regions correspond to the density of solid domains, away from the
interfaces with the pores, and thus are of interest for several reasons.
First, mechanical response properties of porous materials depend on
density of the solid phase. Second, diffusivity (as mentioned above,
it is highly suppressed in the glassy states) is generally a decreasing
function of density~\cite{r49}. Here, we focus on the behavior of the
average density of solid domains as a function of porosity. To get a
measure of the solid phase average density, we define the quantity
denoted as $\langle\rho\rangle_{S}$, such that it provides an average
over the density distributions in the region around the maxima
corresponding to the solid phases (in Fig.\,\ref{fig4n}, the average
solid phase density corresponds to the peaks around $\langle\rho\rangle
_{R}=1.2$). To find the average we assume that the peaks in regions
close to $\langle\rho\rangle_{R} = 1.2$ can be approximated as Gaussian
forms. Then, the parameters of each Gaussian curve can be obtained
from the corresponding numerical fits. The porosity, $p$, was computed
using the methods and the computer code developed by the authors of
Refs.\,\cite{r46,r47}. In Fig.\,\ref{fig6}, we plot $\langle\rho\rangle
_{S}$, as a function of $p$. As follows from the figure, the solid-phase
density of the porous systems does not differ much from that of the
dense ones. In the wide porosity range studied, the maximum variation
in the density is less than $\simeq5\,\%$. The behavior of $\langle\rho
\rangle_{S}$ with porosity, however, possess some noteworthy features.
First, there is a rather abrupt change in $\langle\rho\rangle_{S}$ in
the region of small values of $p$. Second, in the high porosity limit,
$p \geq 0.6$, the density is a strongly increasing function of $p$. At
the intermediate values of $p$, $\langle\rho\rangle_{S}$ demonstrates
a moderate growth. This range spans $p$ values from $\simeq0.1$ to
$\simeq0.5$. Next, we discuss the behavior of porosity with the average
density, $\rho\sigma^3$. The result is shown in the inset to Fig.\,
\ref{fig6}. As our analysis shows, the behavior can be represented by
the following scaling relation $p \propto(\rho_{c}\sigma^3-\rho\sigma^3
)^{\gamma}$. In this formula, $\rho_{c}\sigma^3$ is the critical density,
corresponding to $p=0$. Curiously enough, the value, obtained from the
fitting is: $\rho_{c}\sigma^3=1.24$. Note that the value $\rho_{c}\sigma
^3=1.25$ was reported in Ref.\,\cite{r45} as the critical density below
which LJ systems develop voids upon isochoric energy minimization. The
computed power-law exponent is $\gamma\simeq2.0$. The above relation
works well at both the small and the intermediate values of $\rho
\sigma^{3}$. At the values of $\rho\sigma^{3}$ greater than $\simeq0.8$,
deviations from the scaling behavior occur. Note that, in this region,
$\langle\rho\rangle_{R}$ is a strongly increasing function of $p$. This
means that the density changes drastically with a minor variation of $p$.
The behavior is consistent with the large negative pressures at this value
of $\rho\sigma^3$.

\section{Conclusions}
\label{sum}

In this paper, we studied the structural and thermodynamic
properties of model glassy systems with varied porosity, obtained by
an isochoric rapid quench from a liquid phase to low-temperature
solid phases. The transition to vitreous phase occurs as an
instability, defined by negative derivatives of the chemical
potential and pressure with respect to the concentration and the
local density, respectively. The equilibrium temperature of the
glassy phases was varied in a relatively wide range to reveal
effects of temperature on the thermodynamical properties and
structural specifics of the porosity patterns. We computed the
temperature dependence of the pressure in the systems with different
densities and deduced a scaling law governing the behavior of the
pressure-temperature with average densities. Further, the pore-size
distribution functions were studied. We found that in the systems
with porosity exceeding a characteristic value, the distribution
function obey a single scaling relation. For the highly dense
systems, the distribution resembles closely a Gaussian; this finding
being in agreement with available experimental data. It was found
that a change in temperature of the solid phase does not alter
general shape of the pore-size distribution functions, the major
effect being a widening of the curves. The local density
distribution functions for samples with varied density were also
scrutinized. We found that local density of the solid phase is a
decreasing function of the porosity in the high porosity limit,
while it decreases with the porosity for dense systems. The present
study can be of use for design of porous absorbent materials and
related technologies. Also, it can provide some insight into
mechanism of glass transition and structural defects in
vitreous systems.

\section{Acknowledgement}
\label{ack}

The authors are indebted to Prof. M. Paluch (Institute of Physics,
University of Silesia) for illuminating comments on the experimental
aspects of the studies of glass transition in confined environments.
Financial support from the NSF (CNS-1531923) is gratefully
acknowledged. Computational work in support of this research was
performed at Michigan State University's High Performance Computing
Facility and the Ohio Supercomputer Center.


\end{document}